# Tuning the Band Structures of a 1D Width-Modulated Magnonic Crystal by a Transverse Magnetic Field

K. Di,[1] H. S. Lim,[1,a)] V. L. Zhang,[1] S. C. Ng,[1] M. H. Kuok,[1] H. T. Nguyen,[2] M. G. Cottam[2]

[1]*Department of Physics, National University of Singapore, Singapore 117542*

[2]*University of Western Ontario, Department of Physics and Astronomy, London, Ontario N6A 3K7, Canada*

Theoretical studies, based on three independent techniques, of the band structure of a one-dimensional width-modulated magnonic crystal under a transverse magnetic field are reported. The band diagram is found to display distinct behaviors when the transverse field is either larger or smaller than a critical value. The widths and center positions of bandgaps exhibit unusual non-monotonic and large field-tunability through tilting the direction of magnetization. Some bandgaps can be *dynamically* switched on and off by simply tuning the strength of such a static field. Finally, the impact of the lowered symmetry of the magnetic ground state on the spin-wave excitation efficiency of an oscillating magnetic field is discussed. Our finding reveals that the magnetization direction plays an important role in tailoring magnonic band structures and hence in the design of dynamic spin-wave switches.

## I. INTRODUCTION

Magnonic crystals[1-7] (MCs), with periodically modulated magnetic properties on the submicron length scale, behave like semiconductors for magnons and hold great promise for applications in emerging areas like magnonics and spintronics. Such artificial crystals exhibit unique properties not present in natural materials, such as magnonic bandgaps, and offer novel possibilities in controlling the propagation properties of the spin waves (SWs) or magnons. Recently, enormous theoretical and experimental efforts have been devoted to studying the fundamental properties and potential applications of MCs. Of all the characteristics, the formation of magnonic band gaps, the most essential feature, is most thoroughly investigated. Studies have shown that the band gaps are controllable by tuning the geometrical dimensions,[1,3,4,8] lattice symmetry,[6,8] or constituent materials[3,9] of the MCs.

The magnetic field dependence of band structures is of particular interest. Both theory and experiment reveal that magnonic band gaps can be monotonically shifted and resized in frequency by simply varying the strength of an external applied magnetic field.[5] The great majority of studies were restricted to cases where the equilibrium magnetization is nearly uniformly aligned along an external field in either the backward-volume or Damon-Eshbach geometries. There is to date, however, very little knowledge of the band gaps of MCs for intermediate-field cases where the magnetization is arbitrarily oriented with respect to the SW wavevector, making the field dependence more complex.

---
a) phylimhs@nus.edu.sg.

Here, employing three independent theoretical methods that give consistent results, we present a detailed study of the SW band structure of a one-dimensional width-modulated MC with a static magnetic field applied *transversely* to the long axis of the waveguide system. The band structure exhibits different behaviors depending on whether the field is below or above a critical value. The widths and center frequencies of magnonic band gaps are shown to be tunable in a non-monotonic fashion by varying the applied field. Interestingly, some bandgaps attain their maximal width at a field below the critical value, when the equilibrium magnetization is in an intermediate configuration. Finally, results for the excitation efficiency of the SWs by an applied oscillating magnetic field are also presented and interpreted using group-theory analysis.

## II. CALCULATION METHODS

The $[P_1, P_2]$ = [9 nm, 9 nm] Permalloy MC studied (period $a$ = 18 nm, thickness of 10 nm and alternating widths of 24 nm and 30 nm) has the same geometry and material property as those of Ref. 4. But now we include a variable transverse magnetic field $H_T$ [see Fig. 1 (a)]. To calculate its SW band structure, three independent numerical calculations are performed, based on different assumptions and models and yet yielding consistent results, namely, a finite-element method, a microscopic approach, and time-domain simulations, as described below. A proper understanding of the bandgap properties entails a complete description of the SW modes. In the calculations, the saturation magnetization $M_S$, exchange constant $A$ and gyromagnetic ratio $\gamma$ of Permalloy were set to 8.6×10$^5$ A/m, 1.3×10$^{-11}$

J/m and $2.21\times10^5$ Hz·m/A, respectively, with magnetocrystalline anisotropy and surface pinning neglected.

The finite-element approach was implemented in COMSOL Multiphysics[10] software with calculations done within one unit cell of the waveguide as illustrated in Fig. 1 (a). First, the system was relaxed to its energy minimum under a transverse magnetic field $H_T$ by solving the Landau-Lifshitz-Gilbert (LLG) equation. The demagnetizing field $\mathbf{H}_d = -\nabla \Psi(\mathbf{r})$ is obtained by solving the following Poisson's equation

$$\nabla^2 \Psi(\mathbf{r}) = \nabla \cdot \mathbf{M}(\mathbf{r}), \quad (1)$$

where $\Psi(\mathbf{r})$ is the scalar potential of the demagnetizing field and $\mathbf{M}(\mathbf{r})$ the total magnetization. For the nonmagnetic domain, $\mathbf{M}(\mathbf{r})$ is set to zero. Application of the periodic boundary conditions $\Psi(x+a) = \Psi(x)$ and $\mathbf{M}(x+a) = \mathbf{M}(x)$ automatically guarantees the periodicity of the exchange field. The band structure was then obtained by solving the three-dimensional linearized Landau-Lifshitz equation,[1]

$$\frac{i\omega}{\gamma}\mathbf{m} = \mathbf{M}_0 \times \left[\mathbf{h}_{dip} + (\nabla \cdot D\nabla)\mathbf{m}\right] + \mathbf{m} \times \left[\mathbf{H}_T + \mathbf{H}_d + (\nabla \cdot D\nabla)\mathbf{M}_0\right], \quad (2)$$

where $\mathbf{m}$ and $\mathbf{M}_0$ are the dynamic and equilibrium magnetizations with three components, $\mathbf{h}_{dip} = -\nabla\psi(\mathbf{r})$ is the dynamic demagnetizing fields, and $D = 2A/\mu_0 M_S^2$. The Bloch-Floquet boundary conditions $\psi(x+a) = \psi(x)\exp(ik_x a)$ and $\mathbf{m}(x+a) = \mathbf{m}(x)\exp(ik_x a)$ are applied, where $k_x$ is the SW wavevector in the $x$ direction, $\psi(x)$ and $\mathbf{m}(x)$ are the respective dynamic components of the demagnetizing field and magnetization. Note that in both steps, the same tetrahedral mesh grid was employed and convergence was obtained for various mesh sizes smaller than the exchange length of Permalloy $l_{ex} = \sqrt{2A/\mu_0 M_S^2} \approx 5$ nm.

In the microscopic approach we employed an extension to the dipole-exchange theory in Ref. 11 that was applied to lateral periodic arrays of ferromagnetic stripes coupled via magnetic dipole-dipole interactions across nonmagnetic spacers. In the waveguide MCs considered here, there are interfaces between the Permalloy regions in adjacent unit cells [see Fig. 1 (a)], and so that inter-cell as well as intra-cell short-range exchange has to be taken into account. This is in addition to the long-range dipole-dipole coupling involving all cells. The waveguide was modeled as an infinitely long spatially-modulated array of effective spins that were arranged on a simple cubic lattice, with the effective lattice constant $a_0$ chosen to be shorter than the Permalloy exchange length $l_{ex}$ mentioned above. The appropriate number of spins is then chosen in any of the physical dimensions of the waveguide unit cell so the correct size is obtained (e.g., in the thickness dimension a choice of $a_0 = 1$ nm would correspond to ten cells of spins). The spin Hamiltonian $H$ can be expressed as a sum of two terms:

$$H = -\frac{1}{2}\sum_{\mu,\mu',j,j'}\sum_{\alpha,\beta} V^{\alpha,\beta}_{j\mu,j'\mu'} S^\alpha_{j\mu} S^\beta_{j'\mu'} - g\mu_B H_T \sum_{\mu,j} S^y_{j\mu}, \quad (3)$$

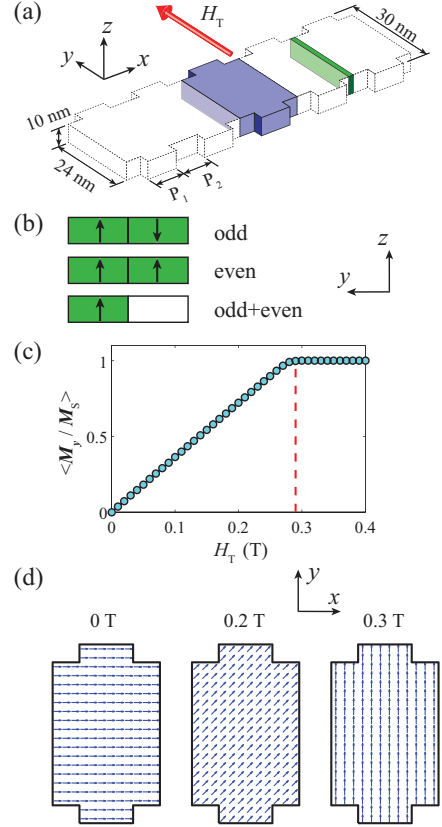

FIG. 1. (a) Schematic view of the magnonic crystal waveguide. The shaded blue block represents the computational unit cell used in the finite-element and microscopic calculations. The green bar indicates the region where the excitation field is applied in the OOMMF simulations. (b) Instantaneous cross-section profile of the excitation field for OOMMF. (c) Average value of the normalized equilibrium transverse magnetization $M_y/M_S$ calculated as a function of the transverse field strength $H_T$. (d) Ground state magnetization of one unit cell subjected to various $H_T$.

where $V^{\alpha,\beta}_{j\mu,j'\mu'}$ is the total (exchange plus dipolar) interaction between the spin components $S^\alpha_{j\mu}$ and $S^\beta_{j'\mu'}$, and the summations are over all distinct magnetic sites, labeled by an index $j$ (or $j'$), which specifies the position within any unit cell, and $\mu$ (or $\mu'$), which labels the repeated cells of the MC along the $x$ direction. The second term in Eq. (3) represents the Zeeman energy due to the transverse field $H_T$ in the $y$ direction, where $g$ and $\mu_B$ denote the Landé factor and Bohr magneton, respectively. The interaction (where $\alpha$ and $\beta$ denote the Cartesian components $x, y$ or $z$) has the form

$$V^{\alpha,\beta}_{j\mu,j'\mu'} = J_{j\mu,j'\mu'}\delta_{\alpha,\beta} - g^2\mu_B^2 \frac{|\mathbf{r}_{j\mu,j'\mu'}|^2 \delta_{\alpha,\beta} - 3r^\alpha_{j\mu,j'\mu'}r^\beta_{j\mu,j'\mu'}}{|\mathbf{r}_{j\mu,j'\mu'}|^5}, \quad (4)$$

where the first term describes the short-range exchange interaction $J_{j\mu,j'\mu'}$, assumed to have a constant value $J$ for nearest neighbors and zero otherwise. The second term describes the long-range dipole-dipole interactions, where $r_{j\mu,j'\mu'} = (x_j - x_{j'} + (\mu - \mu')T, y_j - y_{j'}, z_j - z_{j'})$ is the separation between spins. The parameters of the microscopic and macroscopic models (see Refs. 11,12) are related by $M_S = g\mu_B S/a_0^3$ and $D = SJa_0^2/g\mu_B$, where $a_0$ is the effective lattice parameter mentioned above, $S$ is the spin quantum number, and $g\mu_B = \gamma/2\pi$ in terms of the gyromagnetic ratio.

The steps in the microscopic theory involve first solving for the equilibrium spin configurations in the array, using an energy minimization procedure appropriate for low temperatures and treating the spins as classical vectors. Next the Hamiltonian $H$ was re-expressed in terms of a set of boson operators, which are defined relative to the *local* equilibrium coordinates of each spin. Finally, keeping only the terms up to quadratic order in an operator expansion, we solve for the dipole-exchange SWs of the waveguide MC. In general, the procedure requires numerically diagonalizing a $2N \times 2N$ matrix, where $N$ is the total number of effective spins in any unit cell (one period) of the MC. For most of our numerical calculations we employed values for $a_0$ in the approximate range 1.0 to 1.4 nm, which implies $N \sim 1800$ or larger. The coupled modes will depend, in general, on the Bloch wavevector component $k_x$ associated with the periodicity of the MC in the $x$ direction. By means of a Green's function approach,[12] the microscopic theory can also be used to calculate a mean-square amplitude for the SW modes. This can be done for each of the discrete SW bands as a function of wavevector $k_x$ and the position anywhere in the unit cell for one period of the MC.

In the time-domain calculation of the spin dynamics using OOMMF,[13] the waveguide extending 4-μm in the $x$ direction was discretized to individual $\Delta x \Delta y \Delta z = 1.5 \times 1.5 \times 10$ nm$^3$ cuboid cells. The equilibrium magnetization of the waveguide subject to a transverse field $H_T$ was obtained through solving the LLG equation with a relatively large Gilbert damping coefficient $\alpha = 0.5$. Next a pulsed magnetic field $H(t) = H_0 \text{sinc}(2\pi f_0 t)$ in the $z$ direction, chosen to excite *all* SW branches, was applied to a $1.5 \times 15 \times 10$ nm$^3$ central section of the waveguide [see the odd+even field of Fig. 1 (b)]. This contrasts with the previous two methods where no excitation field is needed. The time evolution of the magnetization was obtained with the damping coefficient $\alpha$ set to 0.005. Dispersion relations are then acquired by performing a Fourier transform of the out-of-plane component $m_z$ in time and space, with contributions from all the discretized magnetizations considered.

### III. MAGNONIC BAND STRUCTURES UNDER TRANSVERSE FIELDS

Under an increasing transverse field $H_T$ applied in the $y$ direction, the equilibrium magnetization in the waveguide is gradually oriented from the $x$ to the $y$ direction due to the competing demagnetizing and applied field. The simulated average $M_y/M_S$ versus $H_T$ plot presented in Fig. 1 (c), indicates that the waveguide is almost completely magnetized in the $y$ direction above a critical field of about $H_C = 0.29$ T. This value accords well with the analytically estimated average demagnetizing field of 0.32 T, based on the assumption that the infinitely-long waveguide has a uniform average width of 27 nm and is uniformly magnetized in the transverse ($y$) direction. For all values of $H_T$ considered, the magnetization is nearly uniform, since the exchange interaction is dominant due to the relatively small size of the waveguide.

The calculated dispersions of magnons in the MC waveguide under transverse magnetic field $H_T = 0$, 0.2 and 0.3 T are shown in Fig. 2. It is clear that, for all $H_T$ values, the MC waveguide exhibits complete magnonic band gaps arising from Bragg reflection and anti-crossing between counter-propagating modes. The frequencies of each branch, at the $\Gamma$ and $X$ points, as functions of field strength are presented in Figs. 3 (a) and (b), respectively. At the $\Gamma$ point, instead of being a monotonic function, the frequencies of all five branches first decrease for $H_T < H_C$ and then increase for $H_T > H_C$, where $H_C \approx 0.29$ T. Such a behavior under a transverse magnetic field has been experimentally observed in ferromagnetic nanowires[14,15] of rectangular or circular cross-section, for which the SW wavevector component along the wire axis is zero. Fig. 3(b) shows that at the $X$ point, the curves feature an obvious dip only for branches labeled 2 and 4, and the fields at which a frequency minimum occurs are lower than $H_C$.

Of special interest is how the complete magnonic band gaps change with magnetization as $H_T$ is varied. Interestingly, Figs. 3(c) and (d) reveal that the overall variation of the bandgap parameters with increasing field is non-monotonic, which contrasts with previous reports of MCs.[1,5-7] With increasing field, the width of the first band gap first increases and then decreases. It remains constant after the equilibrium magnetization is totally aligned along the $y$ direction. We note that a maximum value of ~ 7.1 GHz is attained at about $H_T = 0.2$ T, when the average angle between $M_0$ and the $x$ axis is about $\pi/4$. By comparison, the second band gap decreases monotonically and vanishes above $H_T = 0.25$ T. The center frequency of the first (second) band gap monotonically increases (decreases) as the field increases.

It should be noted that such band gap tunability is mainly a result of the field-tunable separation between center frequencies of gap openings induced by coupling between counter-propagating modes,[4] instead of the tunable widths of the gap openings. We illustrate this using the first band gap (see Fig. 4). In Ref. 17, it is reported that without a static magnetic field, only even-symmetry branches 1, 3, and 5 (odd-symmetry branches 2 and 4) will be excited by an even (odd) excitation field, meaning that there is no coupling between the even and odd modes. Therefore, gap openings appears between bands with the same symmetry, e.g., between 1 and 3 or 2 and 4. When $H_T$ = 0 T, the gap opening is ~ 9.7 GHz between branches 1 and 3, and ~ 21.5 GHz between branches 2 and 4, the frequency overlap of which gives the first complete band gap of ~ 2.8 GHz. It is clear from Fig. 4 (b) that the relatively small change in the widths of the gap openings cannot explain the large variation of the first complete band gap width for $H_T < 0.2$ T. Fig. 4 (c) indicates that for $H_T < 0.2$ T, the center frequency of the first (second) gap opening shifts up (down) sharply, leading to a decreasing center-to-center distance of the gap openings, which should be the main reason for the observed band gap tunability. A similar discussion shows that when $H_T$ increases, the two gap openings responsible for the second complete band gap shift away from each other, causing the band gap to decrease and finally to close fully.

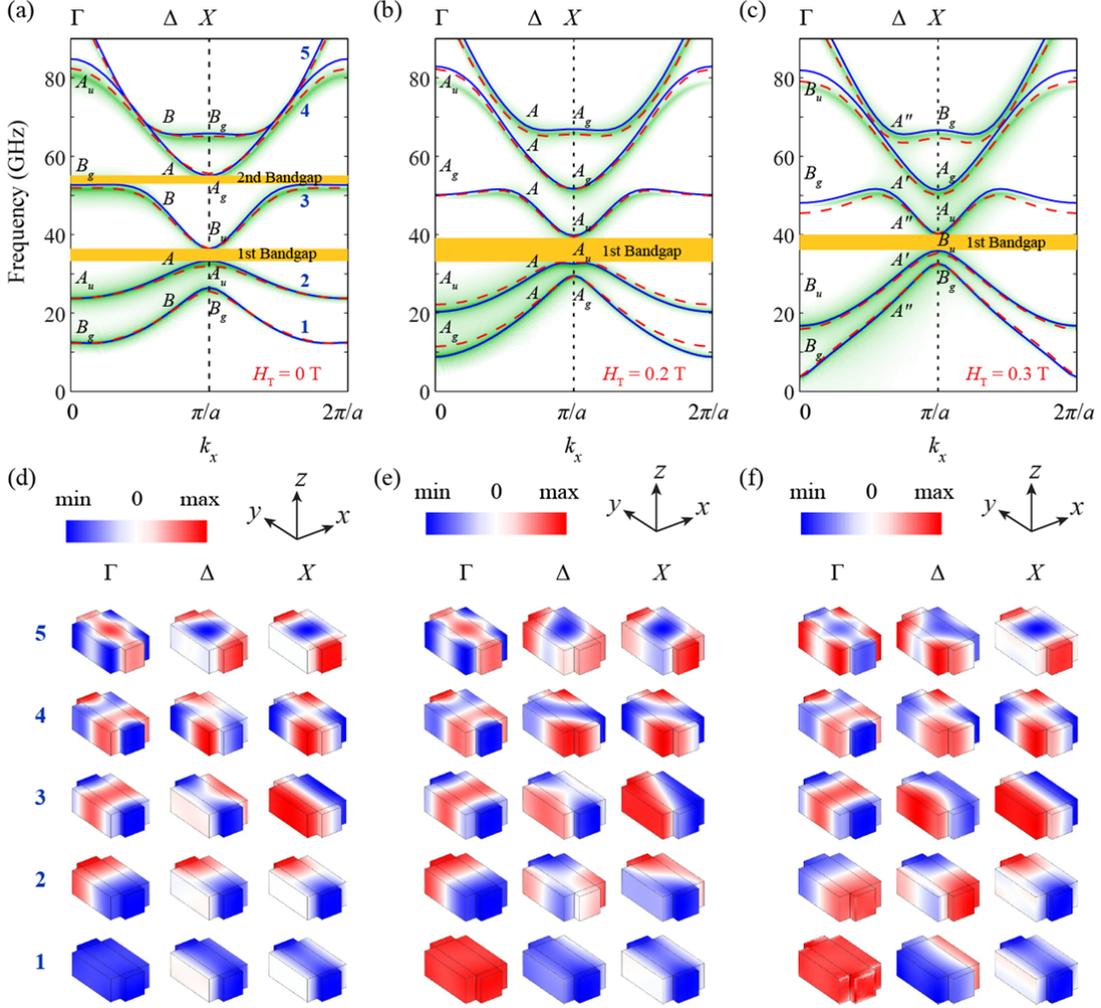

FIG. 2. (a – c) Magnonic dispersion curves under various transverse magnetic fields. The blue solid, red dashed and green bold lines are data obtained from COMSOL, microscopic and OOMMF calculations, respectively. The wavevector at the Δ point is ≈ 0.13 nm$^{-1}$. (d – f) COMSOL-simulated mode profiles of $m_z$ for $H_T$ = 0, 0.2 and 0.3 T, respectively.

## IV. MODE SYMMETRY AND EXCITATION EFFICIENCY

The magnon modes, calculated using the three methods with consistent results, can be classified based on their mode profile symmetry of the dynamic magnetization, as indicated in Fig. 2. The respective symmetries of the magnetic ground state [see Fig. 1 (d)] for $H_T$ = 0, 0.2, and 0.3 T correspond to the respective $C_{2h}$, $C_i$, $C_{2h}$ groups, whose character tables [16] are presented in Table 1. The lower symmetry associated with the ground state for $0 < H_T < H_C$ precludes the labeling of the branches as even or odd symmetry. For $H_T$ = 0 or $H_T > H_C$, the ground state has a higher symmetry as the magnetization is completely saturated along either the $x$ or $y$ direction, respectively. The symmetry group of the wavevector at the Γ and X points is the same as that of the ground state, while that at a general point Δ is usually a subgroup, leading to a lower symmetry. For $H_T$ = 0, 0.2, 0.3 T, the symmetry groups corresponding to the wavevector at Δ are $C_2$, $C_1$, and $C_{1h}$, respectively. In all the cases considered, the axis of rotational symmetry, if one exists, is coincident with the direction of the transverse field.

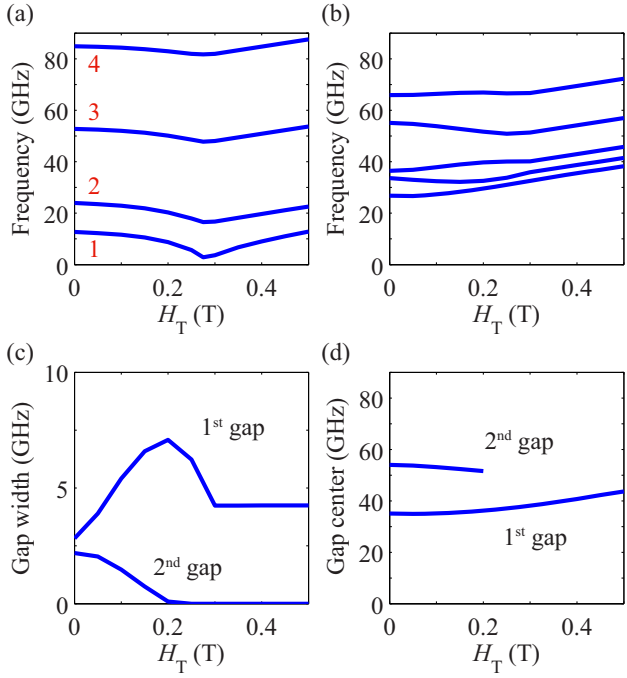

| $C_{2h}$ | $E$ | $C_2$ | $\sigma_h$ | $i$ |
|---|---|---|---|---|
| $A_g$ | 1 | 1 | 1 | 1 |
| $A_u$ | 1 | 1 | -1 | -1 |
| $B_g$ | 1 | -1 | -1 | 1 |
| $B_u$ | 1 | -1 | 1 | -1 |
| $C_2$ | $E$ | | $C_2$ | |
| $A$ | 1 | | 1 | |
| $B$ | 1 | | -1 | |
| $C_i$ | $E$ | | $i$ | |
| $A_g$ | 1 | | 1 | |
| $A_u$ | 1 | | -1 | |
| $C_{1h}$ | $E$ | | $\sigma_h$ | |
| $A'$ | 1 | | 1 | |
| $A''$ | 1 | | -1 | |
| $C_1$ | $E$ | | | |
| $A$ | 1 | | | |

TABLE 1. Character tables of symmetry groups (after [16])

FIG. 3. The transverse field dependences of (a) mode frequencies at the Γ point, (b) mode frequencies at the $X$ point, (c) complete band gap widths, and (d) center frequencies of band gaps.

Using time-domain OOMMF simulations, we have earlier established that[17], for the same MC in the absence of an external field, only $A$ ($B$) symmetry branches can be excited by an $A$ ($B$) symmetry magnetic field [corresponding to odd (even) field in Fig. 1 (b)] for time-domain OOMMF simulations. However, for a transverse field below $H_C$, the above conclusion becomes inapplicable as no point symmetry operation (besides the identity operation) exists for modes at a general point Δ in the Brillouin zone. This is illustrated by the absence of symmetry for the corresponding mode profiles. Fig. 5 indicates that, although the amplitude of $m_z$ at the Δ point has inversion symmetry, its phase, on the other hand, has no such symmetry. In the linear excitation regime, the excitation efficiency for a particular mode $m(r, t)$ by an excitation field $h(r, t)$ is proportional to the overlap integral[18] of $m(r, t)$ and the torque exerted by $h(r, t)$:

$$\eta \propto \langle m, M_0 \times h \rangle = \int m^* \cdot (M_0 \times h) dV = M_0 \cdot \int h \times m^* dV, \quad (5)$$

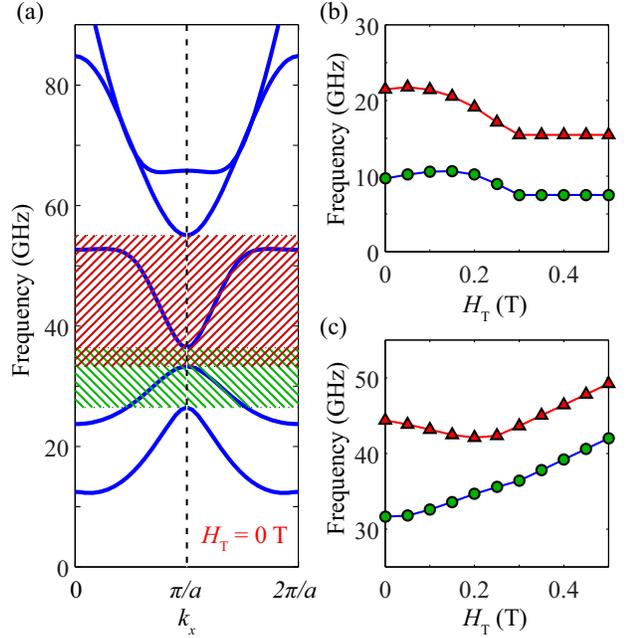

FIG. 4. (a) Formation of the first complete band gap as the frequency overlap between the first (green hatched area) and second (red hatched area) gap openings. (b) Widths and (c) center frequencies of the first (green circle line) and second (red triangle line) gap openings.

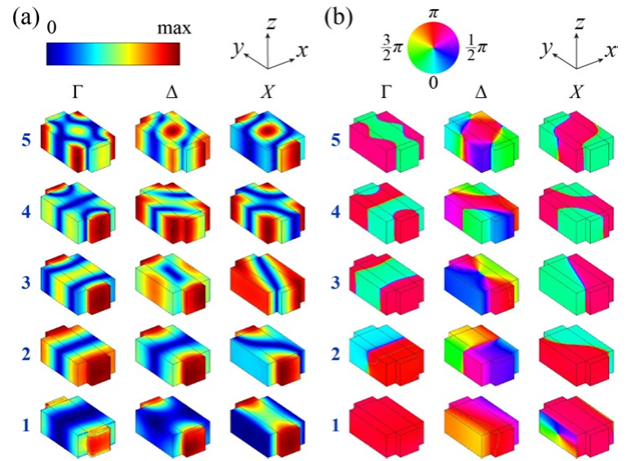

FIG. 5. The (a) absolute values and (b) phases of the dynamic magnetization $m_z$ for the respective modes, at the Γ, Δ and $X$ points, of the five dispersion branches in Fig. 2. The mode profiles were calculated for $H_T = 0.2$ T.

where $M_0$ is regarded as effectively uniform since the lateral size of the MC lies within the exchange-dominated regime, $m^*$ the complex conjugate of $m$, and the integration over volume $V$ is done within regions where the excitation field is nonzero.

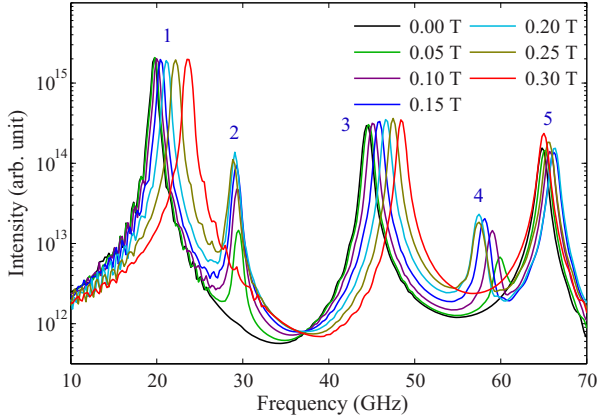

FIG. 6. Magnon spectra, at the $\Delta$ point, excited by the same even-symmetry field and calculated for various $H_T$ values. Groups of peaks corresponding to the five branches are labeled 1, 2, 3, 4 and 5 respectively.

For the regimes of $H_T = 0$ and $H_T > H_C$, only modes with the same symmetry as that of the excitation field have non-vanishing excitation efficiency. In contrast, in the $0 < H_T < H_C$ range, there is no simple correspondence between the symmetry of a mode and that of the excitation field. In this last case, an excitation field with an even distribution [middle panel of Fig. 1 (b)] in the $y$-$z$ plane can excite modes of all branches (i.e. of both even and odd symmetries). Figure 6 presents the OOMMF-simulated excitation spectra excited by the same even excitation field for $H_T$ ranging from 0 to 0.3 T. The modes of branches 2 and 4 can only be excited for $0 < H_T < H_C$, with the maximum efficiency occurring at $H_T \approx 0.2$ T. It is to be noted that Lee et al.[19] claimed that the $A + B$ field [corresponding to the odd + even field in Fig. 1 (b)] in Ref. 17 is not sufficiently general to generate complete magnonic band structure. However, based on Eq. (5) and micromagnetic simulations, we have shown that the $A + B$ field can indeed excite all the modes. Additionally, this is true even for the single-width nanostripe considered in Ref. 19, because the demagnetizing-field-induced effective pinning[20] of the dynamic magnetization along the $y$ axis results in a relatively small yet nonzero excitation efficiency by the $A + B$ field. Although the relative excitation efficiency generally decreases with increasing node number in the $y$-direction, our simulations show that the high-order $m = 3$ mode can be excited by the $A + B$ field, which can be identified under logarithmic scale (not shown). Finally, for the OOMMF simulations, using just one layer of cells across the thickness provides sufficient accuracy for the frequency range considered. For higher frequencies, the above discussion can be trivially extended to include perpendicular standing spin waves (PSSW).

## V. DISCUSSION AND CONCLUSION

Our dynamic magnetic-field tunability of the bandgap has advantages over that based on structural dimensions or material composition. For instance, it is not feasible to reshape band structures in the latter case by tuning corresponding parameters once the MCs are fabricated. In contrast, we have demonstrated here that bandgaps can be dynamically tuned by applying a transverse field. More importantly, by changing the direction of magnetization, some bandgaps can be reversibly switched on and off. Chumak et al. showed that a periodically applied magnetic field opens bandgaps in uniform spin-wave waveguides termed dynamic magnonic crystals.[21] We have shown, conversely, the possibility to dynamically close a bandgap by employing a simple uniform magnetic field, which may be utilized for nanoscale SW switches.

In conclusion, the band structure of a 1D width-modulated nanostripe MC under a transverse magnetic field has been studied using three independent theoretical approaches, the size and center frequency of magnonic bandgaps are found to be highly tunable by the transverse field. Furthermore, some bandgaps can be *dynamically* switched on and off by simply varying the field intensity, providing novel functionalities in magnonics. Further analysis shows that the bandgap tunability arises from the tunable separation between gap openings instead of the width of the gap openings. The breaking and recovering of the ground state symmetry due to the transverse magnetic fields are shown to have important implications for the mode classification and excitation efficiency of an excitation field. We have analyzed the role of the excitation field, which is inherent in the OOMMF simulation method. A full description of the magnonic band structure is obtained, consistent with the other two methods which we emphasize do not involve any choice of excitation field. Also, we have shown that, contrary to the assertion of Lee et al.,[19] the $A + B$ field is able to excite all modes of the MC. Possible applications of our transverse-field results are the excitation of magnonic modes having an odd number of nodes across the stripe width, and dynamically tunable SW switches and filters.

## ACKNOWLEDGEMENTS

This project is supported by the Ministry of Education, Singapore, under Grant No. R144-000-282-112, and the Natural Sciences and Engineering Research Council (Canada).